# Color similarity study among small galaxy groups members

*Ahmed M. Fouad[a,b]\*, Z. Awad[b], A. A. Shaker[a], Z. M. Hayman[b]*

[a] *National Research Institute of Astronomy and Geophysics (NRIAG), 11421, Helwan, Cairo, Egypt.*

[b] *Astronomy, Space Science, and Meteorology Department, Faculty of Science, Cairo University, 12613 Giza, Egypt.*

## A B S T R A C T

We applied a membership test based on the color similarity of group members to detect the discordant galaxies in small groups (quintets) that had been determined by the Friends-of-Friends (FoF) algorithm. Our method depends on the similarity of the color indices (u-g) and (g-r) of the group members. The chosen sample of quintets was extracted from "Flux- and volume-limited groups for SDSS galaxies" catalog which is a spectroscopic sample of galaxies originally taken from the Sloan Digital Sky Survey – Data Release 10 (SDSS-DR10). The sample included 282 quintets with a total number of 1410 galaxies. The similarity measure used in this study is the Euclidean distance. The calculations showed that 73.4% of the group samples (207 out of 282 quintet groups) have galaxies with similar colors (u-g) and (g-r). Each of the remainder groups (75 systems) has an interloper galaxy with different colors than the other members, and hence they became quadrants. We found that group members tend to be more luminous than outliers. We conclude that using the similarities in the color indices between group members gives better identification of group membership.

## 1. Introduction

Galaxies are not randomly distributed in space. They tend to gather in gravitationally bound systems throughout their formation and evolution, and therefore about half of them are found in groups and clusters (Karachentsev 2005). The study of the relation between galaxy properties and their group environment is very important for understanding the evolution of galaxies (Samir et al. 2016; Shaker et al. 1998). On the large-scale, groups and clusters of galaxies are parts of the galactic filaments; hence they drive the structure formation in the Universe. The definition of these systems extends from binaries to rich clusters and superclusters. In 1877, the first group of galaxies, known as Stephan quintet, was observed by Edouard Stephan (Stephan 1877). The initial systematic searches for clusters, ( e.g. Abell 1958; Hickson 1982; Rose 1977; Zwicky et al. 1961), used criteria based on visual identification of galaxy densities on the sky. Thereafter, large catalogs of the groups were constructed from redshift surveys contained more than 1000 groups,( e.g. Giuricin et al. 2000; Merchan et al. 2000; Ramella et al. 2002; Tucker et al. 2000). Recently, large redshift surveys have yielded an accurate measure of galaxy distances. Therefore, several studies benefit from this advantage and applied automated algorithms on such surveys (or galaxy samples) producing more numerous groups and clusters catalogs in the 3-dimensional space,(e.g. Berlind et al. 2006; Eke et al. 2004; Robotham et al. 2011; Tempel et al. 2014; Yang et al. 2007).

The problem with these studies is that the properties of these groups depend on the group finder algorithms that are, in turn, based on the observed redshift of galaxies as a line-of-sight distance **measure. This measure** suffers from uncertainty because the peculiar motions of galaxies distort the line-of-sight structures. Thus, the distinction between real groups and both galaxies within other looser groups or chance alignment field ones is very difficult.

Many studies discussed the clustering dependence on galaxy properties. For instance, the galaxy color dependence studies are presented in many sources such as (Hermit et al. 1996; Loveday et al. 1995; Willmer et al. 1998). Later studies (Li et al. 2006; Madgwick et al. 2003; Norberg et al. 2002; Zehavi et al. 2005, 2002) applied on larger surveys of 2dF Galaxy Redshift Survey (2dFGRS) and the Sloan Digital Sky Survey (SDSS). At redshift ($0.2 < z < 1$), other surveys were conducted to investigate the color dependency of galaxy clustering, (e.g. Carlberg et al. 1997, 2001; Firth et al. 2002; Phleps et al. 2006; Shepherd et al. 2001).

In addition, the most common techniques that have been used to identify group members based on their colors are: the maxBCG technique (Annis et al. 1999), the cut-and-enhance (CE) method (Goto et al. 2002) which selects the members that are similar in colors, and the four-color clustering (C4) algorithm (Nichol et al. 2003). The C4 algorithm was developed by Nichol et al. (2003) to differentiate between the cluster-like and the filed-like galaxies. It is based on previously defined properties of the field measured from a large sky survey. This algorithm exploits the quality and quantity of multi–dimensional astronomical datasets such as the SDSS. It defines galaxy clusters as an overdensity of galaxies in both space (angular position and redshift) and the rest–frame of four colors in order to minimize the contamination due to projection. (Miller et al. 2005) applied the C4 algorithm to the second data release of SDSS and presented the "C4 Cluster Catalog" which contains about 2500 clusters and a new sample of 748 clusters of galaxies is identified. Recently, two machine learning algorithms were used to identify galaxy groups and clusters based on galaxy color similarities (Mahmoud et al. 2016, 2018).

This study aims to identify interloper galaxies of small groups of five members (quintets) based on their color dissimilarities to the other members of the galaxy group by using a distance measure technique following (Mohamed & Fouad 2017; Sabry et al. 2012) with a specific selection criteria.

The layout of this paper is as follows; in Section 2, we present the galaxy group sample used in this study. We also describe the updating process of the observational parameters of all galaxies. In Section 3, we outline the methodology used in detecting group membership and identifying the interloper galaxies. Section 4 describes the results of our work with a discussion. The conclusions are summarized in Section 5.

## 2. Galaxy group sample

In this section, we give a brief description of the catalog used in selecting our sample. We used the SDSS-DR14 dataset of the quintet galaxy groups chosen from the catalog of galaxy groups and clusters (Tempel et al. 2014) which is available online at (http://vizier.cfa.harvard.edu/viz-bin/VizieR?-source=J/A+A/566/A1).

(Tempel et al. 2014) restricted their study to the spectroscopic sample obtained from the Catalog Archive Server (CAS2) of the SDSS-DR10 (York et al. 2000). To construct flux-limited and volume-limited galaxy group and cluster, Tempel et al. used a modified friends-of-friends (FoF) method with a variable linking length in two directions. They took into account the dynamical mass estimates of galaxies in groups depending on the measured radial velocities and group extent in the sky.

The flux-limited catalog includes 588,193 galaxies and 82,458 groups down to apparent magnitude in r-band, $m_r$ = 17.77 mag. The volume-limited catalogs are complete for absolute magnitudes in r-band down to $M_{r, lim}$ = -18.0, -18.5, -19.0, -19.5, -20.0, -20.5, and -21.0; the completeness is achieved within different spatial volumes. The original data covers a field size of 7221 square degrees representing 17.5% of the full sky.

The authors assumed the Wilkinson Microwave Anisotropy Probe (WMAP) cosmological parameters: the Hubble constant $H_0$ = 100 h km s$^{-1}$ Mpc$^{-1}$, the matter density $\Omega_m$ = 0.27 and the dark energy density $\Omega_\Lambda$ = 0.73 (Komatsu et al. 2011).

(Tempel et al. 2014) visually checked and cleaned the data from the spurious entries of incorrect luminosities. Then, they filtered the galaxies to include only the Galactic extinction corrected r-band magnitude galaxies $m_r \leq$ 17.77. Redshifts were corrected for the motion relative to the Cosmic Microwave Background (CMB) and set the upper distance limit to redshift z = 0.2. They calculated k-corrections with the KCORRECT (v4_2) algorithm following (Blanton & Roweis 2007). After applying their method (FoF-algorithm), the final flux- and volume-limited catalogs and tables are constructed and summarized (see their Table 1, (Tempel et al. 2014). They included individual galaxies with richness (number of group member galaxies) equals 1 and groups with richness $\geq$ 2.

In the present study, we utilize the volume-limited catalog of $M_r$ = -18 and maximum redshift of z= 0.045 (hereafter, Mrlim18) which we summarized its data in Table 1.

Table 1: A summary of the observed groups in Mrlim18 showing the total number of galaxies in each group category and their percentage to the whole number of observed galaxies in the catalog.

| Group category | Number of galaxies | Percentage % |
|---|---|---|
| Individual | 20050 | 40.2 |
| Pairs | 7936 | 15.9 |
| **Quintets** | **1845** | **3.7** |
| Other | 20029 | 40.2 |
| Total | 49860 | 100 |

This study focuses on quintets that include only 3.7% of galaxies in Mrlim18. Knowing the galaxies positions from SDSS-DR10, we updated the quintets' data from SDSS-DR14 (Blanton et al. 2017) using the Catalog Archive Server Jobs System (CasJobs) facility that is available online through the SDSS website at (http://skyserver.sdss.org/CasJobs/). We set the search radius (nearest primary object) to 2 arcseconds. Then we extracted the following parameters:

1- The SDSS magnitudes (u, g, r, i and z) that include reddening corrections at the position of each galaxy from the SDSS table entitled (GALAXY) in which corrections are computed using (Schlegel et al. 1998) methodology.

2- The photometric redshift, its estimate error and the rest frame absolute magnitude in the r band from the "Photoz" table.

3- The spectroscopic redshift, its error and the spectroscopic class of the object (GALAXY, QSO, or STAR) from table called "SpecObj".

Tempel et al. (2014) had filtered the galaxy sample and checked visually some galaxies to assure that the sample does not contain stars, quasars and any other spurious entries. However, we found some differences **in the object's** class between SDSS-DR10 and -DR14. Hence, we first re-filtered the chosen sample by removing galaxies that were re-classified in the SDSS-DR14 as stars or quasars from their new spectra. In addition, we excluded galaxies without spectroscopic or photometric redshifts, and galaxies with large errors (>10%) in the *ugriz* bands. In each step of filtration when an object is excluded, we removed its galaxy group from the original list of quintets in the original Mrlim18 catalog. Our filtration process of the selected sample of quintet groups ended up with 1410 galaxies in 282 groups instead of 1845 galaxies in 369 groups.

## 3. Method and Criteria

We define the outlier galaxy as the galaxy that has different color than the other galaxy members in a given group according to (Sabry et al. 2012) criterion. Our method is applied to detect these outlier galaxies in the studied sample of groups. We used a distance measure to define the distances in the color-color diagram between the galaxy members in a given group. The distance measure used is the "Euclidean distance coefficient" (EDC), denoted here by ($e_{ij}$) as expressed in Eq. (1) following (Mohamed & Fouad 2017; Sabry et al. 2012).

$$e_{ij} = \left[ \sum_{k=1}^{2} \left( x_{ik} - x_{jk} \right)^2 \right]^{1/2} \tag{1}$$

where the two subscripts i & j are the galaxy luminosity rank of the studied group as described by (Tempel et al. 2014) while . $x_{ik}$, k=1, 2 represents the color indices of the i[th] member of the group, respectively.

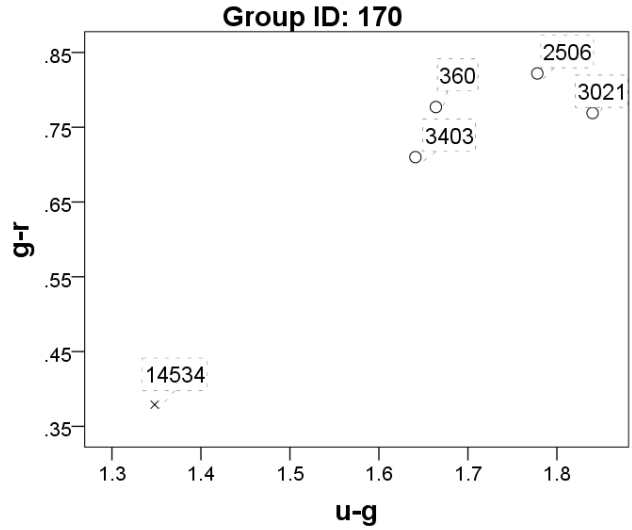

Figure 1. The distribution of galaxies of group (ID: 170) in the **color-color diagram** (u-g) and (g-r).

The coefficient ($e_{ij}$) quantifies the dissimilarity between the i$^{th}$ and j$^{th}$ galaxies. The larger the e$_{ij}$, the dissimilar the galaxies are.

Figure 1 represents the color-color diagram for quintet group ID: 170, taken from(Tempel et al. 2014), which we are going to use as an example to explain, in details, our methodology. From Fig.(1), it is clear that the location of the galaxy ID: 14534 is far from those of the other members of the group. This indicates dissimilarity in its color which, in turn, implies that the galaxy is an outlier. Calculations of the coefficients e$_{ij}$ (Eq.1) is needed to confirm this finding. The calculated coefficients are arranged in a dissimilarity matrix (see Table 2) which is symmetric with respect to the diagonal. Therefore, only 10 entries either above or below the diagonal are used to calculate the average ($e_{av}$) and the standard deviation ($\sigma$) of the coefficients.

Given the $e_{ij}$, $e_{av}$ and $\sigma$, we categorized the group members into four categories according to the following condition;

$$0 < e_{ij} \leq e_{av} - \sigma \qquad \text{Twins (T) (very similar).}$$
$$e_{av} - \sigma < e_{ij} \leq e_{av} \qquad \text{Pairs (P).}$$
$$e_{av} < e_{ij} \leq e_{av} + \sigma \qquad \text{Members (M).}$$
$$e_{ij} > e_{av} + \sigma \qquad \text{Attribute Discordant (AD) (very dissimilar).}$$

For the large dataset used in this study, 1845 galaxies, we defined an arbitrary weight for each of the above categories in which Twin =1 while AD = 4. For the i$^{th}$ member, if its $\sum_{j=1}^{5}(e_{ij}) \geq 12$ then it is considered as an outlier. This occurs for members that has categories either (three AD) or (two M and two AD) in their categories.

**Table 2:** A Summary of the calculations and Mrlim18 data of the group member 170. Col.1. member ID; Col.2: tow color indices; Col.3: is the luminosity rank; Col.4: is the calculated dissimilarity matrix while Col. 5 is the categories of the galaxies. The last column defines the membership of each galaxy to the group (M: Member, O: Outlier). For this example, $e_{av} = 0.803$ and $\sigma = 0.222$.

| Gal ID | Color Indices | | Rank | Dissimilarity matrix | | | | | Categories | | | | | Membership |
|--------|------|------|------|---------|---------|---------|---------|---------|---|----|----|----|----|------------|
| | u-g | g-r | | 1 | 2 | 3 | 4 | 5 | | | | | | |
| 360 | 1.66385 | 0.77723 | 1 | 0 | 0.12234 | 0.1768 | 0.07139 | 0.50865 | 0 | P | P | T | M | M |
| 2506 | 1.7777 | 0.82202 | 2 | 0.12234 | 0 | 0.08232 | 0.17688 | 0.61755 | P | 0 | P | P | AD | M |
| 3021 | 1.84045 | 0.76874 | 3 | 0.1768 | 0.08232 | 0 | 0.20785 | 0.62839 | P | P | 0 | P | AD | M |
| 3403 | 1.64121 | 0.70952 | 4 | 0.07139 | 0.17688 | 0.20785 | 0 | 0.4422 | T | P | P | 0 | M | M |
| **14534** | **1.34771** | **0.37876** | **5** | **0.50865** | **0.61755** | **0.62839** | **0.4422** | **0** | **M** | **AD** | **AD** | **M** | **0** | **O** |

## 4. Results and Discussion

We applied the statistical criteria discussed in Section 3 on the galaxy sample extracted from SDSS-DR14 for 282 quintets to identify the discordant galaxies. The number of groups that has discordant galaxies is 75 systems (27% of the sample after the re- filtration step explained in Section 2). Table 3 lists the excluded (outlier) galaxy IDs, coordinates and their corresponding group IDs, as well. Excluding these members changed the group categories from quintets to quartets (with 4 members). From the above discussion, we conclude that the studied sample is, in fact, 75 quartets and 207 quintets.

In Table 2, the example of group (ID: 170); the galaxy (ID: 14534) is detected as an outlier because it has a total weight, $\sum_{j=1}^{5} (e_{5j}) = 14$. The SDSS-images of this group (illustrated in Figure 2) supports our calculations and confirms (visually) that the excluded galaxy has a different color compared to the other four galaxies.

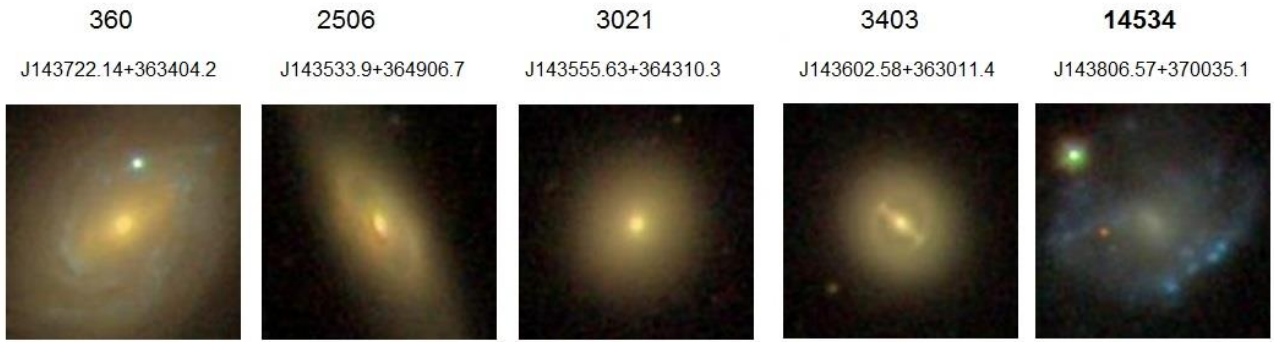

Figure 2. The SDSS color images of group (ID: 170) galaxies. Showing the color difference between the first four members and the outlier galaxy (ID: 14534). All images have size of ≈ 1 x 1 arcminutes.

**Table 3**: A summary of the excluded 75 galaxies from our original sample, taken from Tempel et al. (2014). The table lists the excluded galaxy ID, its group ID, and the galaxy position (coordinates) as given in the SDSS-DR10.

| Galaxy ID | Group ID | Galaxy position | | Galaxy ID | Group ID | Galaxy position | |
|---|---|---|---|---|---|---|---|
| | | RA (α) deg | Dec (δ) deg | | | RA (α) deg | Dec (δ) deg |
| 15280 | 56 | 181.041 | 1.826 | 208096 | 2428 | 183.516 | 13.573 |
| 14534 | 170 | 219.527 | 37.010 | 34836 | 2483 | 245.287 | 13.128 |
| 45553 | 195 | 241.945 | 23.791 | 348362 | 2497 | 184.871 | 60.923 |
| 14889 | 506 | 235.551 | 23.800 | 63355 | 2579 | 171.240 | 34.576 |
| 56991 | 596 | 152.471 | 14.730 | 8554 | 2667 | 218.363 | 9.245 |
| 18192 | 635 | 215.101 | 17.760 | 125422 | 2902 | 224.492 | 1.474 |
| 1415 | 672 | 128.422 | 54.549 | 127929 | 2925 | 122.207 | 5.711 |
| 38089 | 699 | 154.434 | 16.810 | 9396 | 2951 | 221.360 | 19.466 |
| 37737 | 709 | 153.507 | 14.777 | 89858 | 3008 | 125.255 | 38.861 |
| 57936 | 728 | 199.019 | 6.377 | 157584 | 3153 | 156.153 | 42.025 |
| 81287 | 738 | 168.108 | 9.567 | 15457 | 3377 | 233.583 | 9.423 |
| 63439 | 783 | 150.838 | 37.197 | 12469 | 3479 | 168.435 | 57.153 |
| 155315 | 794 | 242.696 | 41.984 | 12703 | 3514 | 170.835 | 34.661 |
| 2511 | 810 | 148.132 | 15.775 | 15928 | 3657 | 218.284 | 58.592 |
| 206966 | 941 | 256.875 | 30.232 | 152059 | 3692 | 233.944 | 21.935 |
| 2331 | 1051 | 186.355 | 16.124 | 255207 | 3760 | 200.948 | 55.177 |
| 97397 | 1106 | 251.210 | 38.960 | 32196 | 3765 | 217.798 | 36.303 |
| 66724 | 1133 | 200.771 | 26.855 | 40203 | 3954 | 258.107 | 28.860 |
| 8339 | 1174 | 218.694 | 3.342 | 157725 | 4001 | 234.389 | 49.515 |
| 3272 | 1274 | 205.659 | 24.465 | 51520 | 4101 | 229.221 | 49.096 |
| 181427 | 1279 | 149.771 | 36.848 | 87351 | 4195 | 133.576 | 20.485 |
| 2992 | 1291 | 169.712 | 7.521 | 117519 | 4318 | 208.282 | 7.664 |
| 154949 | 1317 | 209.822 | 59.297 | 20077 | 4320 | 210.920 | 24.677 |
| 5916 | 1362 | 228.401 | 8.086 | 311800 | 4434 | 198.988 | 7.284 |
| 18012 | 1464 | 178.577 | 23.086 | 23671 | 4612 | 156.335 | 5.930 |
| 3749 | 1567 | 152.145 | 12.555 | 24248 | 4649 | 207.778 | 16.600 |
| 3819 | 1583 | 228.773 | 43.151 | 25009 | 4708 | 215.468 | 28.469 |
| 185034 | 1596 | 181.265 | 6.184 | 101073 | 4946 | 227.226 | 5.355 |
| 68881 | 1612 | 125.939 | 11.671 | 85470 | 5065 | 130.768 | 56.289 |
| 3925 | 1617 | 223.686 | 37.413 | 31338 | 5094 | 215.542 | 25.097 |
| 4282 | 1729 | 236.508 | 5.137 | 405468 | 5112 | 151.503 | -0.943 |
| 88138 | 1817 | 153.220 | 4.741 | 32704 | 5155 | 135.001 | 44.757 |
| 4950 | 1948 | 225.679 | 37.949 | 33658 | 5213 | 238.415 | 4.537 |
| 235378 | 1987 | 146.012 | 0.766 | 86015 | 5316 | 184.680 | 11.726 |
| 236750 | 2112 | 251.887 | 22.995 | 70578 | 6198 | 119.072 | 19.150 |
| 14606 | 2254 | 232.949 | 9.475 | 78925 | 6317 | 238.887 | 7.381 |
| 218558 | 2351 | 226.520 | 46.370 | 472540 | 6399 | 124.862 | 15.940 |
| 6703 | 2395 | 227.419 | 1.386 | | | | |

A comparative study by (Deng et al. 2008) showed that isolated (non-member) galaxies tend to be fainter than member galaxies of groups. Therefore, we investigated the luminosity of the outlier galaxies detected in our calculations as isolated galaxies in their host group fields. The Group members were previously ranked within their group according to the absolute magnitude in r-band as given by (Tempel et al. 2014). The most luminous galaxy in the group has a rank equals to 1, fainter members will have higher rank values where the faintest galaxy in the group has a rank of 5. We found that the 75 outliers have high fraction of rank (5) and the fraction decreases with rank as shown in Figure 3. This trend confirms that these outliers do not belong to their assumed groups as they are fainter than the other group members.

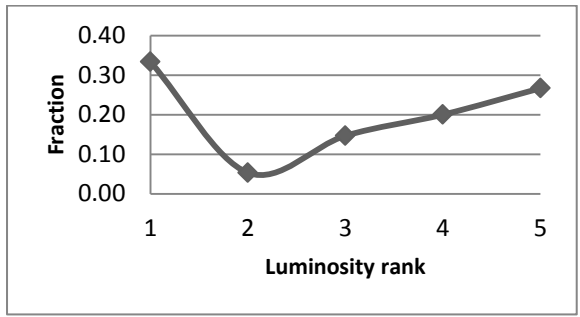

Figure 3. The fraction of outliers according to the luminosity rank following (Tempel et al. 2014).

The exception here is the fraction of the first ranked galaxies (most luminous galaxies), *hereafter* r1. These 25 r1-outliers might be foreground interlopers that are misidentified as members. Based on this assumption, an r1-outlier is expected to have a minimum radial distance compared to the group members. So we investigated the location of these r1 outliers in the groups according to their redshifts. Because of the distortions in redshift space, we don't expect to find them all having minimum redshifts (z_min) but may have high fraction at z_min.

To achieve this, we arranged the 5 galaxies in each group according to the spectroscopic redshift. Each galaxy has closeness notation (c) which ranges from the nearest c1 to the farthest c5 (from zs_min to zs_max). Figure 4 shows the distribution of r1-galaxies according to the closeness for two samples. The first one (solid lines) is the sample of confirmed r1-galaxies of 257 groups (galaxies of all groups (282) excepting the 25 r1-outlier galaxies). The second sample (dashed lines) is the distribution of r1-outliers outliers in the 25 excluded groups. The distribution of the first sample is uniform but the second illustrates higher fractions at c1, c2 and then it decreases with c. Thus, the 25 r1-outliers did not follow the behavior of the other r1-galaxies which confirms our assumption that these outliers are predominantly foreground isolated galaxies.

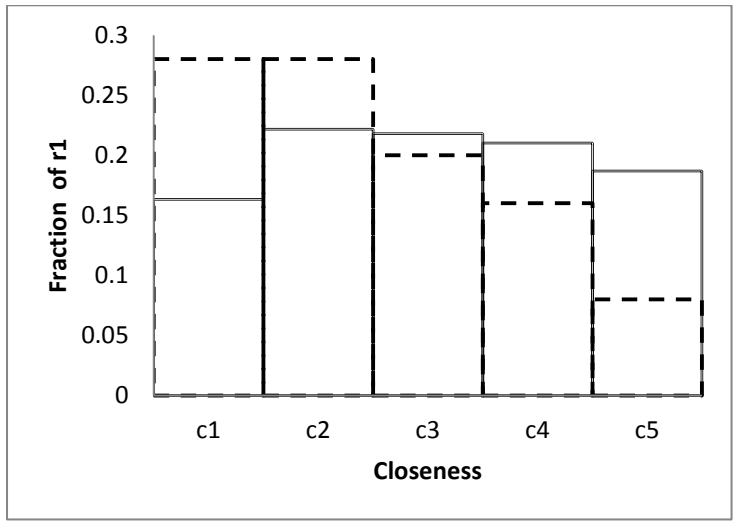

Figure 4. The fraction of the confirmed 257 r1-galaxies (solid lines) and the 25 r1-outliers (dashed lines).

## 5. Summary and conclusions

Galaxy groups are gravitationally bound systems, where the galaxies are orbiting around a common **center**. There are many methods for detecting galaxy groups. Some of these methods rely on the galaxy densities, such as Hickson criteria and FoF algorithm. Other methods depend on galaxy colors (e.g. maxBCG, CE and C4) to reduce contamination by sky projections. We consider the two ideas by studying the color similarity **among** the members of galaxy groups sample identified by FoF algorithm. Our method is based on the Euclidean distance similarity measure. We selected the quintet groups (282 systems) from the volume-limited groups' catalog of $M_r$ = -18 constructed by (Tempel et al. 2014). We found that the groups that host galaxies of similar colors (u-g) and (g-r) are 207 groups which represent 73.4% of the total number of groups. In the reminder groups (75 systems), our method detects an interloper galaxy in each group as it has different colors from the other members. By investigating the common properties of these interloper galaxies, we found that they have a higher proportion of faint galaxies. Finally, we conclude that considering the color similarity between the members of FoF galaxy groups is beneficial to eliminate the contamination by chance alignment galaxies.

## Acknowledgement

The authors would like to thank Dr. Ali Takey from the National Research Institute of Astronomy and geophysics (NRIAG) for his fruitful discussions and comments on this work. This research has made use of the VizieR catalog access tool, CDS, Strasbourg, France. The original description of the VizieR service was published in A&AS 143, 23.

Funding for the Sloan Digital Sky Survey IV has been provided by the Alfred P. Sloan Foundation, the U.S. Department of Energy Office of Science, and the Participating Institutions. SDSS acknowledges support and resources from the Center for High-Performance Computing at the University of Utah. The SDSS web site is www.sdss.org.